\documentclass{aipproc}
\layoutstyle{6x9}

\begin{document}
\title[Measurement for fields]{Models of measurement for quantum fields
           and for classical continuous random fields\footnote{This article
will be submitted to the Proceedings of the Conference on the Foundations of Probability
and Physics-4, V\"axj\"o, 2006.
If it is published, it will be found at http://proceedings.aip.org/proceedings.}
}

\begin{abstract}
A quantum field model for an experiment describes thermal fluctuations
explicitly and quantum fluctuations implicitly, whereas a comparable continuous
random field model would describe both thermal and quantum fluctuations explicitly.
An ideal classical measurement does not affect the results of later measurements,
in contrast to ideal quantum measurements, but we can describe the consequences
of the thermal and quantum fluctuations of classically non-ideal measurement
apparatuses explicitly. Some details of continuous random fields and of Bell
inequalities for random fields will be discussed.
\end{abstract}
\author{Peter Morgan}{address={Yale University},,altaddress={peter.w.morgan@yale.edu}}

\maketitle

\newcommand\Half{{\frac{1}{2}}}
\newcommand\Intd{{\mathrm{d}}}
\newcommand\eqN{{\,\stackrel{\mathrm{N}}{=}\,}}
\newcommand\PP[1]{{(\hspace{-.27em}(#1)\hspace{-.27em})}}
\newcommand\PPs[1]{{(\hspace{-.4em}(#1)\hspace{-.4em})}}
\newcommand\kT{{{\mathsf{k_B}} T}}
\newcommand{\Chi}{{\mathsf{X}}}
\newcommand\SUB[1]{{\kern-.25ex\lower .35ex\hbox{$\scriptstyle #1$}}}
\newcommand\SUBf{{\kern-.25ex\lower .35ex\hbox{$\scriptstyle f$}}}
\newcommand\SUBSUBf{{\kern-.15ex\lower .25ex\hbox{$\scriptscriptstyle f$}}}
\newcommand\MarginAnnot[1]{{\footnote{#1}}}

\section{Introduction}
Ten years ago, I wondered what the differences, similarities, and the relationship,
if any, might be between \emph{quantum fields} and \emph{continuous random fields}.
I had seen that equilibrium states of continuous random fields at non-zero
temperature have nonlocal correlations, as quantum fields do, and I thought
na\"\i vely that this was enough to let us construct useful models.
Until three years ago I called continuous random fields by the name that
seemed natural to me, "classical statistical fields", but I discovered then
that mathematicians and physicists are accustomed to using the name "continuous random
fields", and I now use that name.

I still consider that continuous random fields give an effective way to understand
quantum field theory, but they are not a very effective replacement for quantum field theory.
The relationship between quantum fields and continuous random fields particularly
depends on the significantly different measurement algebras of the two theories,
which I discuss below.
Thus I do not refer to continuous random fields as a \emph{sub}-quantum theory,
which I think indicates a stronger relationship to quantum field theory than can be
sustained in the face of the measurement theory differences.

The next two sections discuss some basics of continuous random fields and the
derivation of Bell inequalities.
The assumptions that have to be made to derive Bell inequalities are generally not
satisfied by random fields.
More detail may be found in \cite{MorganA} and \cite{MorganB}.
The later sections on measurement further develop this way to understand
quantum field theory.

\section{A brief description of continuous random fields}
\label{CONT}
A random field is a very general object, no more than an indexed set of random
variables\cite{Vanmarcke}, which means that random fields subsume almost all
probabilistic physics.
If the index set has additional structure, which in physics is particularly
likely to be a space-time structure, then that additional structure will
be inherited by the random field.
A continuous random field can be defined as a random-variable-valued tempered
distribution,
\begin{equation}
  \chi_\SUBf=\int\chi(x)f(x)\Intd^4x.
\end{equation}
Very similarly, a quantum field is an operator-valued tempered distribution,
\begin{equation}
  \hat\phi_\SUBf=\int\hat\phi(x)f(x)\Intd^4x.
\end{equation}
Both the quantum field and the continuous random field are linear maps from a
Schwartz space of functions\cite[\S II.1.2]{Haag}, into a $\star$-algebra of linear
operators and a space of random variables respectively.
The definition given here for a continuous random field is not the only one possible, for
example see \cite{Rozanov}, but here the intention is to see how closely a continuous
random field can parallel a quantum field, so of course we will follow the quantum
field structure closely when possible.
A classical random field can be formulated even more closely in parallel with quantum
field theory as a commutative quantum field.
A lattice as an index set might also be useful, but will not be used here.

The measurement represented by $\hat\phi_\SUBf$ in quantum field theory is compatible
with $\hat\phi_g$ when the functions $f$ and $g$ have space-like separated
supports\footnote{From now on functions will be taken from a Schwartz space of functions.},
whereas the measurement represented by $\chi_\SUBf$ is always compatible with
$\chi_\SUB{g}$, no matter what the space-time relationships between $f$ and
$g$.\footnote{Note that a continuous random field is \emph{not} a
$\mathrm{C}^\infty$ classical field.
In (very) heuristic terms, a continuous random field that has non-trivial
fluctuations can be thought of at a point as almost always $\pm\infty$
(so definitely not a regular continuous field!), but arranged just so that
when we take a weighted average over a finite region, where the
weight function is taken from a Schwartz space, we get a finite number.
A respectable definition of course does not mention infinity.}

The vacuum sector of the quantized Klein-Gordon field can be presented using just the
following two equations:
\begin{eqnarray}
  &&\left[\hat\phi_\SUBf,\hat\phi_g\right]=(g,f)-(f,g),\\
  &&\left<0\right|e^{i\lambda\hat\phi_f}\left|0\right>=e^{-\,\Half\lambda^2(f,f)},
\end{eqnarray}
where $(g,f)$ is a Hermitian inner product:
\begin{eqnarray}
  (g,f)&=&\hbar\int\frac{\Intd^4k}{(2\pi)^4}\:
               2\pi\delta(k^\mu k_\mu-m^2)\theta(k_0)\:\tilde g^*(k) \tilde f(k)\cr
       &=&\hbar\int\frac{\Intd^3\mathbf{k}}{(2\pi)^3}
         \frac{\tilde g^*(\mathbf{k})\tilde f(\mathbf{k})}{2\sqrt{\mathbf{k}^2+m^2}}.
\end{eqnarray}
The first equation fixes the algebraic structure of the quantized Klein-Gordon
field, while the second fixes the vacuum state.
Taken with the linear structure
$\hat\phi_\SUB{(\lambda f+\mu g)}=\lambda\hat\phi_\SUB{f} + \mu\hat\phi_\SUB{g}$,
they are enough to generate the Wightman functions.
I keep to free fields because renormalization complicates matters in ways that
I believe not to be essential, notwithstanding the usual physicist's cry that
they want to see explicit calculations for an interacting quantum field theory.
All that is needed is any way to generate the Wightman functions for an interacting
theory, which can be any Lorentz invariant formalism, preferably better defined
than a singular Lagrangian formalism (it is not to be forgotten that there is
no rigorous interacting model of the Wightman axioms, which makes the Wightman axioms
open to question).

The vacuum sector of an "equivalent" random field can be presented as:
\begin{equation}
  \left<e^{i\lambda\chi_\SUBSUBf}\right>_0=e^{-\,\Half\lambda^2(f,f)}
\end{equation}
in which all measurements are mutually compatible.
This uses the same exponential that is used for the quantized Klein-Gordon field
as the characteristic function of a classical probability density.
The generated joint probability density for a set of observables
$\,\left\{\chi_\SUB{f_1},\chi_\SUB{f_2},...,\chi_\SUB{f_n}\right\}\,$
is the same as for the corresponding observables
$\ \left\{\hat\phi_\SUB{f_1},\hat\phi_\SUB{f_2},...,\hat\phi_\SUB{f_n}\right\}\ $
if the quantum observables are mutually compatible --- that is, essentially, 
if the quantum observables have space-like separated supports.
For non-vacuum states we generally cannot simply adopt the state over the quantized
Klein-Gordon algebra for an "equivalent" continuous random field, but
positive-definite Wigner functions are empirically adequate for any system that we
observe using only mutually compatible measurements, and if we include enough of the
experimental apparatus this can give an adequate classical model for an experiment.

It is interesting to compare the Wigner function of the vacuum state of the quantized
Klein-Gordon field with the equilibrium state of the classical Klein-Gordon field at finite temperature.
The Wigner function at time $t$ is the inverse functional fourier transform of
\begin{equation}
  \left<0\right|e^{i\hat\phi_{f_t}}\left|0\right>
   =\left<0\right|\exp{\left[i\int\hat\phi(t,\mathbf{x})f_t(\mathbf{x})
                                  \Intd^3\mathbf{x}\right]}\left|0\right>
            =\exp{\left[-\hbar\int\frac{\tilde f_t^*(\mathbf{k})\tilde f_t(\mathbf{k})}
                                       {4\sqrt{\mathbf{k}^2+m^2}}
                                  \frac{\Intd^3\mathbf{k}}{(2\pi)^3}\right]};
\end{equation}
we use $f_t(x)$ and $\Phi_t(x)$ as the transform variables of this Gaussian exponential
to obtain
\begin{equation}
  \rho_0[\Phi_t]\eqN \exp{\left[-\frac{1}{\hbar}\int
         \tilde\Phi_t^*(\mathbf{k})\sqrt{\mathbf{k}^2+m^2}\,\tilde\Phi_t(\mathbf{k})
         \frac{\Intd^3\mathbf{k}}{(2\pi)^3}\right]},
\end{equation}
where $\eqN$ denotes equality up to (infinite) normalization and we take $\Phi_t(x)$ to
be complex as a simple way to extend the configuration space to the phase space.
From a classical perspective, the square root here is nonlocal, in the sense that
a classical dynamics would have to be nonlocal for the Gibbs probability density
to take this exponential form.
The nonlocality drops off exponentially, however, making the square root classically
no more problematic than the heat equation --- it's a Lorentz invariant heat kernel.
From a quantum field perspective this analysis is not pertinent; there is no
nonlocality because we cannot send messages faster than light using the measurements
that are available within the Klein-Gordon quantum field theory, and the introduction
of the concept of a classical Gibbs probability distribution is not consistent with
the general principles of quantum field theory.
Even if we restrict ourselves to concepts of quantum field theory, however, Planck's
constant controls the amplitude of quantum fluctuations, just as $\kT$ controls the
amplitude of thermal fluctuations, so that we can properly talk about and distinguish
between quantum and thermal fluctuations.

The equilibrium state of the classical Klein-Gordon random field at temperature $T$ is:
\begin{equation}
  \rho_C[\Chi_t]\eqN e^{-\beta H[\Chi_t]}
                =\exp{\left[-\frac{1}{\kT}\int
         \tilde\Chi_t^*(\mathbf{k}){\textstyle\Half}{\left(\mathbf{k}^2+m^2\right)}
                      \tilde\Chi_t(\mathbf{k})
         \frac{\Intd^3\mathbf{k}}{(2\pi)^3}\right]}.
\end{equation}
There are three obvious changes from the quantum case:
(1) there's no square root;
(2) Planck's constant becomes $\kT$ (with energy units instead of action units);
(3) the Lorentz symmetry is broken (the classical
Klein-Gordon dynamics is Lorentz invariant, but the
Gibbs probability density as an initial condition is not;
for a general Lorentz invariant dynamics, those for which
the Gibbs probability density is Lorentz invariant are singled out).

We can also compute the Wigner function of the equilibrium state of the
quantized Klein-Gordon field at temperature $T$ from the characteristic function
\begin{equation}
  \mathrm{Tr}\left[e^{-\beta \hat H}e^{i\hat\phi_{f_t}}\right],\qquad
    \hat H=\int a^\dagger(\mathbf{k})a(\mathbf{k})\left(\mathbf{k}^2+m^2\right)
    \frac{\Intd^3\mathbf{k}}{(2\pi)^3},
\end{equation}
leading to
\begin{equation}
  \rho_T[\Phi_t]\eqN
     \exp{\left[-\frac{1}{\hbar}\int\frac{\Intd^3\mathbf{k}}{(2\pi)^3}
   \mathrm{tanh}\!\left(\frac{\hbar\sqrt{\mathbf{k}^2+m^2}}{2\kT}\right)
   \tilde\Phi_t^*(\mathbf{k})\sqrt{\mathbf{k}^2+m^2}\,\tilde\Phi_t(\mathbf{k})\right]}
\end{equation}
The tanh factor makes the integrand close to the classical thermal state at low wave
numbers (where tanh is close to linear), and close to the vacuum state at high wave
numbers (where tanh is close to 1).

We can therefore distinguish precisely between quantum fluctuations and thermal
fluctuations --- using purely conventional quantum field computation --- in a way
that allows us to model quantum and thermal fluctuations classically in the same
way as we already model thermal fluctuations.
We can achieve as much empirical adequacy with a classical model insofar as quantum
phenomena are caused by quantum and thermal fluctuations rather than by our contingent
measurements.

\section{Bell inequalities for random fields}
\label{BELL}
I argue at length in \cite{MorganB} that Bell inequalities cannot in general be
derived for random fields, so that the violation of Bell inequalities by experiment
does not rule out random field models.

Bell describes an experiment statistically using random variables $A$ and $a$
representing measurement results and measurement settings associated with a
space-time region $\mathcal{R}_A$ and similarly in a space-like separated space-time
region $\mathcal{R}_B$ \cite[Chapter 7]{Bell}, resulting in an experimentally
established joint probability $p(A,a,B,b)$.
Of course we only actually have statistics from an experiment, but we generally take
probability densities to be effective models for statistics.
The probability density $p(A,a,B,b)$ is no more than an initial condition for a
random field model, which (partially) determines what the initial conditions of the
random field must have been in the past and will be in the future.
Initial conditions are totally unconstrained by classical physics: if they are
unlikely they have higher free energy, but then they just need more experimental
effort to set up.
Experiments that violate Bell inequalities are quite hard to set up --- they take a
lot of free energy --- so it is somewhat disingenuous of Bell, and many others since,
to say pejoratively that it requires a "conspiracy" \cite[Chapter 12, p. 103]{Bell}
for classical physics to obtain a violation of the Bell inequalities, insofar as
this suggests that the experimenter had little to do with setting up favourable
conditions for obtaining the experimental result.

A quantum field state presented as a Wigner quasi-probability over phase space at
the time of a measurement determines the Wigner quasi-probability over phase space
at future and past times to precisely the same extent as a classical random field,
so there is exactly as much "conspiracy" in a quantum field model as there is in
a continuous random field model.
We should not conclude that quantum field theory is unreasonable, but that
a continuous random field theory is as reasonable.

To derive Bell inequalities for a random field, Bell introduces \emph{a priori}
constraints on what the initial conditions in the past are allowed to have been.
Bell's constraints are based on a notion of common cause that is well-founded for
a classical particle model, but is completely unmotivated for a random field.
For a classical particle model, two particles come from a single source, which is the
common cause of two correlated events.
For a random field (and for a quantum field), the correlations of the field at the
time of measurement evolve from correlations at earlier times.
There is a \emph{distributed} cause, not a common cause.

Bell also constrains the dynamics to be local, which has generally been seen as the
only way to satisfy Lorentz invariance.
The kernel $\sqrt{\mathbf{k}^2+m^2}$ is therefore forbidden because it's nonlocal,
even though it's a Lorentz invariant, exponentially decreasing nonlocality that a
sensible classical random field theory ought to be comfortable enough with --- or
anyway as comfortable as we are with the quantum field that generates the same
probability density over the phase space.

Random field theory constrained this much is a straw man of a theory.
The violation of Bell inequalities can reasonably (though not without \emph{any}
qualification) be taken to show that there are no simply localized particles that
would justify an assumption that there are pervasive common causes, but the
violation does not rule out a random field model.

An alternative discussion of the Bell inequalities asserts that Bell inequalities are
about consistency in the sense of probability theory and measurement incompatibility,
not about Einstein locality (for example,
\cite{Fine,Accardi,deMuynck,Pitowsky,Aerts,Loubenets}, and see Karl Hess and Walter
Philipp in this volume for a historical discussion of papers in the statistics literature
going back to the 1950's).
If we construct probability distributions $p(A_1,B_1)$, $p(A_2,B_1)$,
$p(A_2,B_2)$, and $p(A_2,B_1)$, where $A_1$ and $A_2$ correspond to two
incompatible measurement settings in $\mathcal{R}_A$ and similarly for
$B_1$ and $B_2$ in $\mathcal{R}_B$, we \emph{cannot} in general construct
a quadrivariate probability distribution $p(A_1,A_2,B_1,B_2)$ that has
the post-selected probability distributions $p(A_1,B_1)$, $p(A_2,B_1)$,
$p(A_2,B_2)$, and $p(A_2,B_1)$ as marginals.
A classical description that has $p(A,a,B,b)$ as a marginal, however, is not ruled out,
because here there is no post-selection on incompatible measurement settings.
That is, classical and quantum measurement algebras are theoretically distinct but
empirically equally adequate formalisms for the description of experiments.

The experimental data idealized as a probability distribution $p(A,a,B,b)$
from a single set of compatible measurements cannot compel us to use a
non-positive Wigner quasi-probability over the phase space of a quantum
field theory as the only empirically adequate quantum state.
Any probability distribution over a larger phase space that has $p(A,a,B,b)$
as its marginal over $A$,$a$,$B$,$b$ is as empirically adequate as any
putative quantum mechanically non-trivial (non-positive) Wigner function.
We can only be forced to use a non-positive Wigner function if we make at least
two incompatible measurements of identical subensembles of an ensemble of
systems (and the measurement results are provably incompatible with a positive
Wigner function).

A random field model is \emph{contextual} insofar as it includes apparatus degrees
of freedom, but this is not contextuality in the usually pejorative sense that
particle properties depend on what apparatus is used.
There are no particles that have precise properties in random field models.
We have to include the measurement apparatus in models whenever necessary because
of nontrivial effects of thermal or quantum fluctuations, but as a practical matter of
calculation we avoid introducing fluctuations whenever possible.

\section{Measurement for quantum fields}
It is well known that there is no consistent measurement theory for
quantum field theory.

Suppose we carry out a number of measurements described by operators
$\hat\mathcal{O}_i$ on an ensemble of systems described by a density
operator $\hat\rho$.
The expected experimental results are given by quantum theory as
$\mathrm{Tr}[\,\hat\rho\hat\mathcal{O}_i\,]$.
For quantum theory to work as an empirical theory, measured systems and measurement
systems must be manifestly separable in practice --- there has to be something that
is described by $\hat\rho$ and something distinct that is described by the
$\hat\mathcal{O}_i$.
A much stronger requirement on quantum theory, which is required if we are to claim
more for quantum field theory than empirical adequacy, is to insist that measured
systems and measurement systems are manifestly and precisely separable
\emph{in principle}.

From this elementary point of view, it is very clear in the formalism of
quantum mechanics that a quantum field state is not changed by using different
ideal measurement devices, in the sense that the expected results of measurements
described by operators $\hat\mathcal{O}_i$ are not
$\mathrm{Tr}[\,\hat\rho_i\hat\mathcal{O}_i\,]$, with a different quantum
field state as well as a different measurement operator for every measurement.
If that were the case, we could never determine the density operators
$\hat\rho_i$ by measurement.

It is a peculiarity of quantum field theory that a measurement apparatus that
implements $\hat\mathcal{O}_i$ is so far idealized as to be completely outside
space-time (an \textbf{external} description of measurement).
The quantum field state $\hat\rho$ describes all space-time, with no room for
a measurement apparatus.

In contrast, from a point of view in which a measurement apparatus is explicitly
modelled as in space-time (an \textbf{internal} description of measurement), the
measurement apparatus is \emph{in principle} not separable from the system it
interacts with, because of the Reeh-Schlieder theorem.
When we discuss subsystems in quantum mechanics we generally use partial trace
operations to construct states and POV measurements from states and measurements
of a larger system (for example, see \cite{BGL}), but in quantum field theory we
instead discuss different states in the same Hilbert space, which effectively model
different whole universes.
There is an in principle contradiction in the logical structure of quantum field theory,
if we want to include both internal and external descriptions of measurement in the
formalism (and quantum theory has always abrogated to itself the right to place
the Heisenberg cut anywhere convenient).
This in principle contradiction does not affect the possibility of using quantum
field theory as an empirically adequate theory, because separation in an effective
theory sense is generally possible, but it should qualify our support for quantum
mechanics as a fundamental theory.

In practice, we model experiments by introducing a quantum field state and
operator descriptions for the measurements, in the hope that such an idealized
model will be empirically adequate.
If we can't achieve empirical adequacy with a simple such model because thermal
fluctuations have a significant effect on experimental results, we can model the
measurement device and the "measured system" \emph{explicitly} using a single
quantum field state, then a more remote measurement device is modelled by
an operator.
This style of modelling using quantum field theory falls far short of the claims
generally made for the universality of quantum field theory.

\section{Measurements for random fields}

Ideal classical measurements do not affect other measurements or
the system they measure.
We routinely reduce the temperature of experimental apparatus to very close to absolute
zero, so that it is plausible to idealize an experiment as being carried out actually
at absolute zero, whereas we cannot reduce quantum fluctuations, Planck's constant,
to an arbitrarily small level.
When thermal fluctuations are significant and cannot be reduced, we have to model the
thermal properties of an experiment explicitly, so we introduce successively more careful
thermal models until we achieve an empirically adequate description.
Equally, when quantum fluctuations are significant, and presumably cannot be reduced to
an arbitrarily small level, then all we have to do in a random field approach is
explicitly model the quantum fluctuations of the measurement device and its interactions
with the measured system.
A random field model that explicitly takes quantum fluctuations into account is not
much more difficult than a quantum field model if thermal fluctuations already have
to be taken into account.

Even if we cannot reduce quantum fluctuations, we can \emph{imagine} what results
we would obtain if we could.
It is after all quite common to think of physical systems from what is essentially a
God's eye view of the world, in which the usual rules of action and reaction are
suspended --- indeed, as we have seen in the previous section, quantum field theory
as much requires this idealization as does classical physics.
Real experiments can be described with better accuracy by taking progressively more
account of actions and reactions between parts of the experiment, including more
apparatus and changing our God's eye view as needed.

We can construct random field models from quantum field models by a very simple procedure.
To do so, construct progressively larger quantum field models, including more and more
experimental apparatus, until a positive-definite Wigner function is empirically adequate.
\hspace{2em}\framebox{\textbf{\emph{That's our random field model.}}}\hspace{2em}
By construction, the dynamics of the random field are the same as the dynamics of the
quantum field, all that can be different are the initial conditions, whether probability
or quasiprobability over the phase space.
This approach to constructing empirically equivalent random field models should be
distinguished from Arthurs-Kelly type interpretations of positive semi-definite Husimi
functions\cite{AK,BCM,BGL}.

\section{Discrete measurements}
If we're to think in terms of random fields, we have to be comfortable with thinking
of discrete events as localized thermodynamic events in the field.
A \textbf{discrete event device} is a metastable thermodynamic state that is very carefully
tuned as far as possible not to transition to its registration state (that is, except for
its dark rate statistics) whenever it is effectively isolated.
Examples of such devices are CCDs, photographic plates, semiconductor devices, and
bubble chambers\footnote{We can get discrete results by observing the spectrum of
incandescent sodium atoms, for example, but this kind of discrete structure does not
immediately demonstrate that the world is classically particulate, whereas it is a
commonplace that tracks in bubble chambers are taken to justify "particle" physics.}.
From a theoretical point of view, truly thermodynamic transitions can happen in a
finite region such as a single crystal of a photographic plate, because there are
effectively infinite degrees of freedom in a finite region of a continuous random
field model.

When a transition to the registration state happens, a feedback process returns the
device to its metastable state as quickly as possible (unless it's a photographic plate;
a bubble chamber is returned to its metastable state cyclically rather than by
feedback).
The dark rate statistics of a discrete event device are generally non-zero for any
interesting degree of sensitivity, but a device becomes far more interesting if when it
is put near various plugged-in and turned-on apparatuses, \emph{different} statistics
for transitions to its registration state are observed.
A change of the environment generally changes the response of a discrete event device,
and we are very careful to engineer as large a change of response as possible for a
given change of the environment.

Discrete transition events are a consequence of \emph{engineered} thermodynamic properties
more than of any discrete structure of the external field.
To a great extent, this attitude to discrete event devices is in tune with Copenhagen-type
interpretations, which routinely enjoin us not to think there is a particle, or
at least that a particle has any properties like position, until a measurement is made,
but the approach taken here goes further:
classical experimental apparatus is effectively localized, but between preparation and
discrete event devices there are no point-like objects at all.
However, the approach taken here also stops short of a Copenhagen-type positivism:
there is \emph{something} in our models, a random field, between the preparation and
discrete event devices, and indeed constituting the whole experiment.

Discrete events are taken here not to represent the arrival of individual particles,
but nonetheless there may be integer invariants of a quantum or random field.
As a classical analogy, the topological equivalence class of a continuous field is
a nonlocal discrete invariant.
Comparably, the superselection sector of a quantum field is a nonlocal discrete invariant.

In this "picture", we can take for a trivial example a loop of paper, with a single twist,
and define the number of particles, one, to be the number of twists in the loop.
This is a topological invariant under continuous deformations of the loop.
Suppose that the dynamics of the loop of paper are such that at zero fluctuations the
twist is effectively quite localized, so that we can pragmatically identify where the
twist is on the loop.
We might then identify the particle as being where the twist is, but we would have
to recognize that the number of particles is defined as the number of twists in the
loop, and where the twist is located can be defined only pragmatically and only if there
are \emph{no} quantum or thermal fluctuations.
Once we introduce quantum and thermal fluctuations and continuous random fields,
however, there is to my knowledge no developed theory of discrete structure for
random fields comparable to topological and superselection sectors for classical
continuous fields and quantum fields.

In three-dimensional space, the simplest candidate for a topologically nontrivial
structure is perhaps the Shankar monopole (\cite{Shankar}; see \cite{Nakahara} for
a graphic presentation), but because there are several significant quantum numbers
associated with superselection sectors of the standard model, the Shankar monopole
can only be a toy model.

To consider a double-slit experiment with electrons, when the preparation device is turned
on, some of the substance of the preparation device "boils off" into the surrounding space,
where the evolution of the field is more-or-less linear, and we can effectively think of
the field in terms of simple classical waves, passing through the slits and interfering
in the classical wave-like way.
All through that process, invariants of the field must be conserved, so that we can
more-or-less legitimately talk of the total number of electrons in the whole apparatus,
but we cannot as legitimately talk of ideas that are not well-defined such as the number
of electrons in the preparation device, in the discrete event device, or in between, nor
of the time when an electron left the preparation device or arrived at a discrete event
device (even if we see a discrete event, which is perhaps better thought of as part of
the device statistics in the presence of a specific preparation device, not as caused
by a single electron).
Insofar as we might reasonably talk of how many electrons there are between the
preparation and the discrete event device, to some pragmatic approximation, we can
generally only talk of them as contained in the whole space.
There are no localized electrons between the preparation and discrete event devices.
Because a discrete event device that has been tuned so that its transition statistics
change in the presence of an electron field has been placed beyond the slits, we
observe discrete events, but because the electron field is different in different
places, partly because of interference, the discrete event statistics are different in
different places.

Note that it is only because \cite{MorganB} shows that random fields do not satisfy
assumptions that would allow us to derive Bell inequalities that it is possible to
argue in these vague terms for such a model.
This model for localized and non-localized particles is certainly vague, but it is not
much more vague than descriptions of measurement that ignore the engineering of
discrete event devices and claims that the standard model describes localized matter
through confinement.

\section{Speculations about gravity}

Because we have been using Fourier transforms extensively, constructing a
"random field gravity" is mathematically no easier than quantum gravity, but
conceptually it may perhaps be advantageous to keep everything classical.

A striking consequence of taking a more classical view of the world is that if we
suppose that quantum fluctuations vary from place to place, it is classically
natural to wonder whether a Poincar\'e disc view might be a useful
analogy\cite[Chapter IV]{Poincare}, which would suggest that a variation of Planck's
constant would be indistinguishable from a variation of the metric.
On this view, it is only because we include variations of the metric in our
models that Planck's constant is a constant.
Zahid Zakir, in a Nelson trajectories approach, comes to essentially this
conclusion\cite{Zakir}.

On a classical random field view, it also makes little sense to quantize gravity
if it is conceptually related to quantum theory as an intimately related
collective phenomenon --- if the light metric is not merely affected by quantum
fluctuations.
A classical analogy with sound suggests that the light metric might be as
irrelevant as a dynamical variable below some critical length as the sound metric
is irrelevant at the atomic scale.
This seems quite different from suggestions that the light metric structure becomes,
for example, foam-like below certain scales, presupposing that the light metric is
well-defined at arbitrarily small scales.

\section{Conclusion}
Bell inequalities are always brought out in response to any challenge to quantum
theory, but they are of very limited application to random fields.
Nonetheless, a random field model is not a replacement for a Hilbert space model, it
is an alternative.
Finite-dimensional Hilbert space models and the Schr\"odinger equation on a
finite-dimensional phase space will and should continue to be used as very effective
models.

Some interpretations of quantum mechanics look a little outrageous (but of course
are not ruled out) if a random field model can be equally empirically adequate,
but the intended point of this research is more that a better understanding of the
relationship between quantum field models and random field models is helpful.

By including enough apparatus in models, we can make a positive-definite Wigner
function empirically adequate, and then interpret it as what we would observe if
we had classically ideal measurement apparatus.
This seemingly trivial approach can only be taken if a very careful understanding
of the relationship between the measurement models of quantum field theory and of
continuous random fields is preserved.

\section{Acknowledgements}
I am very grateful to the organizers for the invitation to speak at Foundations of
Probability and Physics-4 and to Yale University for additional financial assistance.
I am grateful also to Walter Philipp for helpful comments on a draft of this paper.

\end{document}